\documentclass[12pt,preprint]{aastex}

\def\be{\begin{equation}}
\def\ee{\end{equation}}

\newcommand{\msun}{{M}_{\sun}}

\catcode`\@=11 
\def\@versim#1#2{\vcenter{\offinterlineskip
        \ialign{$\m@th#1\hfil##\hfil$\crcr#2\crcr\sim\crcr } }}

\begin{document}

\title{On the Origin of Ultraviolet Emission and the Accretion Model of
Low-luminosity AGNs}

\author{Zhaolong Yu\altaffilmark{1,2}, Feng Yuan\altaffilmark{1}, and
Luis C. Ho\altaffilmark{3}} \altaffiltext{1}{Key Laboratory for
Research in Galaxies and Cosmology, Shanghai Astronomical
Observatory, Chinese Academy of Sciences, 80 Nandan Road, Shanghai
200030, China; zlyu@shao.ac.cn;fyuan@shao.ac.cn}
\altaffiltext{2}{Graduate School of the Chinese Academy of Sciences,
Beijing 100039, China}\altaffiltext{3}{The Observatories of the
Carnegie Institution for Science, 813 Santa Barbara Street,
Pasadena, CA 91101, USA; lho@obs.carnegiescience.edu}

\begin{abstract}

Low-luminosity active galactic nuclei (LLAGNs) are generally
believed to be powered by an inner radiatively inefficient,
advection-dominated accretion flow (ADAF), an outer truncated thin
disk, and a jet.  Maoz (2007) recently challenged this picture based
on the observation that the strength of ultraviolet emission
relative to the X-ray and radio bands does not depart from empirical
trends defined by more luminous sources.  He advocates that AGNs
across all luminosities have essentially the same accretion and
radiative processes, which in luminous sources are described by a
standard optically thick, geometrically thin disk.  We calculate
ADAF models and demonstrate that they can successfully fit the
observed spectral energy distributions of the LLAGNs in Maoz's
sample.  Our model naturally accommodates the radio and X-ray
emission, and the ultraviolet flux is well explained by a
combination of the first-order Compton scattering in the ADAF,
synchrotron emission in the jet, and black body emission in the
truncated thin disk.  It is premature to dismiss the ADAF model for
LLAGNs.  The UV data can be fit equally well using a standard thin
disk, but an additional corona and jet would be required to account
for the X-ray and radio emission.  We argue that there are strong
theoretical reasons to prefer the ADAF model over the thin disk
scenario. We discuss testable predictions that can potentially
discriminate between the two accretion models.

\end{abstract}


\section{Introduction}

Supermassive black holes (BHs) reside in
the nuclei of most large galaxies. The luminosity of the
nuclei of these galaxies spans at least 10 orders of magnitude, from
the most active and luminous AGNs, to less active low-luminosity
AGNs (LLAGNs), down to nearly quiescent systems such as our
Galaxy. It is now well known that they are powered by gas accretion
onto the supermassive BH.

Some of the most important information about the physics of AGNs
comes from their spectral energy distribution (SED). The SEDs of
luminous, unobscured AGNs are relatively straightforward to obtain.
One of their most distinguishing features in the optical and
ultraviolet (UV) bands is the ``big blue bump'', which is
conventionally interpreted as thermal emission arising from a
standard optical thick, geometrically thin accretion disk (Shakura
\& Sunyaev 1973; Shields 1978; Malkan \& Sargent 1982). By contrast,
the intrinsic weakness of LLAGNs poses practical challenges to
obtaining reliable data to construct their SEDs. Nevertheless, by
now a sufficiently large number of LLAGNs have been adequately
studied with multiwavelength observations (e.g., Ho 1999; M~81,
Markoff et al. 2008; M~87, Di Matteo et al. 2003; NGC~3998, Ptak et
al. 2004; NGC~4594, Pellegrini et al. 2003) to demonstrate that that
their SEDs are markedly different from those of luminous AGNs. In
addition to their low luminosity, other distinctive observational
features of LLAGNs include: extremely low Eddington ratios, low
inferred radiative efficiencies, the lack of the big blue bump, the
weakness or lack of reflection features in the X-ray band, the
narrowness of the iron K$\alpha$ line, ubiquity of compact radio
cores or jets, the frequent detection of double-peaked broad Balmer
lines, and prevalence of low-ionization nebular conditions (e.g., Ho
1999, 2009; Ho et al.  2000; Terashima et al. 2002; Ptak et al.
2004; see Ho 2008 for a review). Compared to luminous AGNs, the dim
emission in the UV band relative to the optical and X-ray bands
leads to a steep optical-UV slope and tends to make these sources
systematically ``X-ray loud" (relative to the UV band).

The lack of the big blue bump in LLAGNs indicates that the thin disk
must be absent or truncated. Within the truncation radius, the
accretion flow is believed to be replaced by an advection-dominated
accretion flow (ADAF; Narayan \& Yi 1994, 1995; Abramowicz et al.
1995; see Yuan 2007 for a review of observational evidence for the
ADAF model of LLAGNs)\footnote{In addition to ``ADAF'', another
widely used term is ``RIAF'' (radiatively inefficient accretion
flow; e.g., Quataert 2003; Yuan et al. 2003; Ho 2008). The reason
for inventing this term is perhaps because it was realized that mass
loss in the form of outflows plays an important role in reducing the
radiative efficiency of this kind of accretion flow (see Section 2.2
of the present paper). However, detailed studies indicate that mass
loss is not the dominant factor for the low efficiency. For example,
the total radiative efficiency of Sgr A* is $10^{-6}$. Mass loss
contributes about $10^{-2}$ (since $99\%$ of the gas is lost in the
outflow), while the remaining factor of $10^{-4}$ is attributed to
energy advection of electrons (see Yuan et al. 2003; Yuan 2007).
Although direct electron heating is significant in an ADAF (i.e.,
$\delta \approx 0.5$), most of the energy is stored in the electrons
and advected into the black hole (Fig. 1 in Yuan 2006). For other
LLAGNs, since $R_{\rm out}$ is smaller than that in Sgr A*, the
outflow loss is weaker and energy advection plays a relatively more
important role for the low efficiency of the system.}. Compared to
the standard thin disk, the radiative efficiency of an ADAF is much
lower. This explains why the luminosity of LLAGNs is so low even
though there is a relatively large amount of gas available for
fueling (Ho 2008, 2009). This picture is further strengthened by the
narrowness of the iron K$\alpha$ line, since the broad iron line is
attributed to X-ray fluorescence off of a cold accretion disk
extending to small radii (e.g., Nandra et al. 1997). The frequently
detected double-peaked broad Balmer lines also require that the cool
accretion disk must have a relatively large inner radius (Chen \&
Halpern 1989; Ho et al. 2000). Quataert et al. (1999) modeled the
optical/UV spectra of M~81 and NGC~4579 and argued that they are
better fit with a truncated thin disk than a standard thin disk
extending to the innermost stable circular orbit\footnote{Using
updated data, we reach a different conclusion regarding M~81 later
in this paper.}. Ptak et al. (2004) reached a similar conclusion
based on the lack of any reflection feature in the X-ray spectrum of
NGC~3998. In another well-studied LINER, NGC~1097, we can get the
value of the truncation radius from fitting its double-peaked Balmer
emission-line profile; this value is in good agreement with that
required to fit its SED with a truncated thin disk model (Nemmen et
al. 2006).

Further strong evidence for the ADAF plus truncated thin disk model comes
from the study of the BH X-ray binary XTE J1118+480 in its low,
hard state (Yuan et al. 2005). Except for the difference in BH mass, the
hard state of BH X-ray binaries is widely believed to be
physically similar to the conditions in LLAGNs. This source is special
because it has an exceptional complete multiwaveband SED, especially in
the EUV. To fit the EUV data clearly requires the thin disk to be truncated
at $\sim 300 \, R_{\rm S}$ (Chaty et al. 2003; Yuan et al. 2005). Furthermore,
a quasi-periodic oscillation (QPO) with a frequency of
$\sim 0.1$ Hz was detected  in this source. As suggested by Giannios \&
Spruit (2004), the QPO can be excited by the interaction of the inner hot
accretion flow and an outer thin disk.  The QPOs then result from the basic
$p$-mode oscillations of the inner ADAF with a frequency roughly equal to the
Keplerian frequency at the truncation radius. The Keplerian frequency at
$300 \, R_{\rm S}$ is $\sim 0.22$ Hz, which is roughly consistent with the
observed QPO frequency of $0.1$ Hz.

Another distinguishing characteristic of LLAGNs is that they almost always
contain compact radio cores, whose strength (relative to either the optical
or X-rays) often formally qualifies them as being ``radio-loud.''  This is
true not only of LINERs (Ho 1999, 2002, 2008; Terashima \& Wilson 2003), but
also of many lower luminosity Seyferts (Ho \& Peng 2001).  Generally the radio
emission is too strong to be attributed to an ADAF and is more consistent
with a jet origin (e.g., Ulvestad \& Ho 2001; Anderson et al. 2004).
This trend is again qualitatively consistent with the observed properties
of X-ray binaries in their low state (e.g., Fender \& Belloni 2004).

Maoz (2007; see also Pian et al. 2010) presents a different point of
view. Using observations of 13 LLAGNs, focusing especially on data
from the radio, UV, and X-ray bands, he shows that the UV/X-ray
ratio (parameterized as the two-point spectral index $\alpha_{\rm
ox}$) of LLAGNs is similar to that of high-luminosity Seyfert
galaxies, although LLAGNs generally lie below the extrapolation of
the correlation between $\alpha_{\rm ox}$ and UV luminosity defined
by more luminous systems (e.g., Steffen et al. 2006).  He asserts
that there is no evidence for a change in SED shape in LLAGNs
compared to luminous AGNs. His sample confirms the previous finding
that LLAGNs tend to be radio-loud (defined via the radio/UV ratio),
but he dismisses this as indicative that LLAGNs occupy a physically
distinctive state compared to luminous sources, on the grounds that
the AGN population as a whole shifts to greater radio dominance at
lower luminosity or Eddington ratio (Ho 2002; Sikora et al. 2007).
These two findings lead Maoz to conclude that LLAGNs do not form a
distinct population compared to luminous AGNs and that the accretion
flow in LLAGNs can still be described by a standard thin disk.

We disagree with the above conclusion because of the following
reasons. First of all, it is hard to conclude that all AGNs must
obey the same accretion flow model just because they follow similar
empirical correlations. To date, we do not have a definitive
accretion disk model that can self-consistently explain the origin
of X-ray emission in luminous AGNs, and thus a rigorous explanation
for the $\alpha_{\rm ox}$-luminosity correlation is still lacking.
Secondly, even if we do attempt to constrain accretion models from
these correlations, the fact that LLAGNs lie {\em below} the
extrapolation established in high-luminosity AGNs can be interpreted
as evidence that LLAGNs have a {\em different} accretion mode than
luminous AGNs.  The overall similarity of $\alpha_{\rm ox}$ for AGNs
of different luminosities and the roughly monotonic behavior of the
$\alpha_{\rm ox}$-luminosity correlation do indicate that there is
no dramatic discontinuity between the accretion modes of low- and
high-luminosity sources.  However, we should not be surprised by
this result. Suppose that at some accretion rate luminous AGNs make
a transition to LLAGNs. In the ADAF model of LLAGNs, the UV emission
comes from the sum of the radiations of the truncated thin disk and
the ADAF\footnote{The contribution of the jet can be neglected if
the accretion rate is not too low.}. Just after the transition the
accretion rate of LLAGNs is still relatively high; thus, its
efficiency is similar to that of a thin disk, and the truncation
radius of the thin disk is small (Yuan \& Narayan 2004). In this
case, we do not expect the emission from the thin disk and ADAF to
be that distinctive from that before transition.  The main
properties of the system (e.g., $\alpha_{\rm ox}$) should remain
roughly continuous during the transition. The observed trend of
increasing radio output or jet power toward lower luminosity or
Eddington ratio has no simple explanation within the context of the
standard thin disk model.  We simply remark that a number of
theoretical studies suggest that the physical conditions of an ADAF
may be more conducive to launching jets and outflows (Livio et al.
1999; Meier 2001; Narayan 2005).

The above arguments are still largely qualitative. To put them on a more
quantitative footing, in this paper we calculate the emitted spectrum of ADAF
model and compare it with the observed SEDs of the LLAGNs presented in
Maoz (2007).  The aim is to see whether the ADAF model can provide a
reasonable explanation of the SEDs, especially in the UV band. We will
show that the radio, UV, and X-ray emission of most sources can
be reasonably reproduced in the ADAF model without deliberately
adjusting the model parameters.

\section{SED Modeling}

\subsection{Sample Constructions}

We revisit the SEDs of the LLAGNs in the sample of Maoz (2007), except
NGC~404 and NGC~3486, which, as explained by Maoz, may be significantly
contaminated by emission from a central star cluster.  The sample is
summarized in Table 1.  All of these 11 sources are UV-variable, and the
variable UV flux provides a firm lower limit to the intrinsic AGN UV
emission. All the sources in the sample have UV observations obtained with
the {\it Hubble Space Telescope (HST)} in the F250W and F330W filters using
the ACS/HRC (Maoz et al. 2005; Maoz 2007).
Unresolved UV nuclei were found in all of these sources.  To construct more
complete SEDs, we collected additional UV data points, as well as
data at radio and X-ray bands, from the literature (e.g., Ho 1999;
Di Matteo et al. 2003; Ptak et al. 2004; L. C. Ho, in preparation) or from
NED\footnote{http://nedwww.ipac.caltech.edu/}.  To mitigate contamination
from the host galaxy, we were careful to choose only data observed at the
highest possible angular resolution.  In practice, this restricts our selection
to interferometric data for the radio, {\it HST}\ for the UV, and
{\it Chandra}\ and {\it XMM-Newton}\ for the X-rays.  Where they overlap,
and taking into consideration variability, the new, additional UV data show
good general agreement with the UV data of Maoz et al. (2005).

\subsection{The ADAF-jet model}

We use the coupled ADAF-jet model to fit the SEDs of our sample. We
only briefly describe the model here, the details of which can be
found in Yuan et al. (2005, 2009). The accretion flow consists of
two parts: an ADAF within a ``truncation'' radius $R_{\rm tr}$ and a
truncated thin disk outside of $R_{\rm tr}$. Both observations and
theoretical studies indicate that the value of $R_{\rm tr}$ is a
function of the accretion rate or luminosity (Liu et al. 1999; Meyer
et al. 2000; Yuan \& Narayan 2004). In this work, we set $R_{\rm
tr}$ as a free parameter. It mainly determines the emitted spectrum
of the truncated thin disk, while its effect on the ADAF is very
small since the radiation of ADAF comes from the innermost region of
the accretion flow. We find that the fitted values are roughly
consistent with the correlation shown in Yuan \& Narayan (2004;
their Fig. 3). In the inner region of the accretion flow, a fraction
of the material is assumed to be transferred into the vertical
direction to form a jet. In the ADAF model, we parameterize the
accretion rate with a parameter $s$, defined such that
$\dot{M}=\dot{M}_{\rm out}(R/R_{\rm tr})^{s}$, where $\dot{M}_{\rm
out}$ is the accretion rate at $R_{\rm tr}$. This is because
simulations of both non-radiative (e.g., Igumenshchev \& Abramowicz
1999; Stone et al. 1999) and strongly radiative (Yuan \& Bu 2010)
accretion flows indicate that hot accretion flows are strongly
convectively unstable. As a result, only a fraction of the gas
available at large radius actually accretes onto the BH, and the
rest of the gas is lost into a convective outflow. Other parameters
for calculating the global dynamical solution of ADAFs are the
viscosity parameter $\alpha$, magnetic parameter $\beta$, and
$\delta$, the fraction of the turbulent dissipation that directly
heats the electrons. In most cases, we adopt their ``typical''
values widely used in the literature: $\alpha = 0.3$, $\beta = 0.9$,
$\delta = 0.5$, and $s = 0.3$, except that in a few cases we also
try other values of $\delta$, given the large uncertainty of the
microphysics of direct electron heating (e.g., Sharma et al. 2007).
This assumption is justified because we believe that these
parameters should have the same values in various objects; they
should be determined by the microphysics of the accretion flow or
jet, which is usually independent of the accretion (or loss) rate
and black hole mass. For example, numerical simulations of accretion
flows with significantly different accretion rates show that the
value of $s$ is identical (Stone et al. 1999; Yuan \& Bu 2010).
Holding these microphysics parameters fixed should not affect the
qualitative features of our conclusions.

The radiation model of the jet is based on the internal shock
scenario of gamma-ray burst afterglows (see Yuan et al. 2005 for
details). Compared to the accretion flow, there are more
uncertainties in the jet model. The main parameters include the mass
loss rate $\dot{M}_{\rm jet}$, the half-opening angle $\phi$, the bulk
Lorentz factor $\Gamma_{\rm jet}$, the power-law index $p$ of the
accelerated relativistic electrons in the jet, and the fraction
of the shock energy that enters into the electrons and magnetic
field, $\epsilon_e$ and $\epsilon_B$. Throughout this paper we adopt
``typical'' values of $\phi=0.1$ and $\Gamma_{\rm jet}=10$, so they
are not free parameters. All other parameters are treated to be free, although
some theoretical constraints are expected. For example, shock
acceleration theory typically gives $2 < p < 3$. However, there is
still some uncertainty in our understanding of shock acceleration,
and, more generally, the acceleration mechanism of electrons in
jets. For instance, magnetic reconnection may be another relevant
mechanism in addition to shocks. Thus, we also try values of $p < 2$.
Medvedev (2006) showed that electron acceleration in relativistic shocks
should roughly follow $\epsilon_e \sim \sqrt{\epsilon_B}$.

The three components in our model are coupled, and thus their
parameters are not completely free but related with each other. The
ADAF and the truncated thin disk are connected at $R_{\rm tr}$, and
the accretion rates of the two components at $R_{\rm tr}$ must be
identical (i.e., $\dot{M}_{\rm out}$). The mass loss rate in the jet
must be a reasonable fraction of the mass accretion rate of the ADAF
in the inner region where the jet is launched; there are large
theoretical uncertainties involved, but typical values generally lie
in the range $\sim 1-10\%$. Moreover, the minimum Lorentz factor of
the accelerated electrons in the jet, which is an important
parameter affecting the radiation of the jet, is also partly
constrained by the electron temperature of the ADAF.

The process of our modeling is as follows. We first guess the value
of the two free parameters of the ADAF compoment, $\dot{M}_{\rm
out}$ and $R_{\rm tr}$. Note that these two parameters are not
completely free, as explained above. Combined with the other model
parameters (e.g., black hole mass, $\alpha, \beta$, etc.), we solve
the radiation hydrodynamical equations of an ADAF to get the global
solution. This yields all dynamical quantities of the system, such
as density, temperature, and magnetic field.  We then calculate the
radiative transfer to obtain the emitted spectrum. Using
$\dot{M}_{\rm out}$ and $R_{\rm tr}$, the spectrum emitted from the
truncated thin disk can be easily computed. The radiation of the jet
can be obtained once the values of $\dot{M}_{\rm jet}$, $p$,
$\epsilon_e$, and $\epsilon_B$ are given. From the discussion in the
last paragraph, we know that $\dot{M}_{\rm jet}$ must satisfy some
general constraints. We then sum the radiation from the the ADAF,
the truncated thin disk, and the jet, and compare the result with
the observed multi-waveband spectrum. We adjust the parameters so
that their sum can fit the overall spectrum satisfactorily, ranging from
radio, optical, UV, to X-rays.

We do not perform any rigorous statistical analysis to test the
goodness of fit, nor are fits themselves always unique. The primary
reason is technical.  Because of the great difficulty in obtaining a
transonic global solution of an ADAF, it would be very
time-consuming, if not impossible, to do a complete parameter survey
for every object. Our aim is not to explore the full allowed
parameter space of ADAF-jet-disk models, but rather to find one such
model for each object that is roughly consistent with its data. We
judge the goodness of the fit simply by ``eye.''   For the problem
at hand, we think that it is physically not appropriate to perform a
rigorous comparison between the theoretical prediction and the
observational data. As in most other theoretical models, all three
components of our model have large and hard-to-quantify
uncertainties, since many simplifications and approximations have to
be adopted. Under these circumstances, the results of any rigorous
quantitative statistical analysis will be completely meaningless.
Given the approximate nature of the model, the comparison between
theoretical prediction and data is, by necessity at this point,
semi-subjective.

It is worth emphasizing that this is not an easy task, and the
flexibility of the model is actually not large. For example, for the
sources presented in Figure 1, the X-ray spectrum is dominated by
the ADAF. It is rather challenging to fit both the normalization and
the shape of the spectrum while not violating the data at other
spectral bands using only one parameter ($\dot{M}_{\rm out}$). Thus,
a successful fit of the X-ray spectrum by the ADAF component is a
nontrivial task. For the sources presented in Figure 2, the
contribution of the ADAF to the X-ray spectrum is negligible. If we
increase $\dot{M}_{\rm out}$, the model will overpredict the radio
flux. For the jet component, we have four free parameters. Their
values are well constrained by fitting the radio spectrum, and we
have little room to adjust the contribution of the jet in the
optical, UV, and X-ray bands. Regarding the truncated thin disk
component, we do have some freedom on the value of $R_{\rm tr}$,
although not so much since its value is correlated with
$\dot{M}_{\rm out}$. But the value of $R_{\rm tr}$ only determines
where the spectrum emitted by the truncated thin disk is cut off.

\subsection{Results}

We fit the radio to X-ray spectrum of all sources in the sample
using the coupled ADAF-jet model. The
results are shown in Figures 1--3, and the adopted parameters of
each source are listed in Table 1. The main results are as follows.

First, with the exception of one object (NGC~1052), the model can
successfully fit the SEDs of these sources ranging from the radio to
the hard X-rays.  In particular, the model fits the UV flux
reasonably well. We want to point out that since the values of the
free parameters in our model have almost been fixed when fitting the
spectrum in the radio and X-ray bands, there is little room for us
to adjust them to fit the UV data points. One interesting result is
that for most sources in our sample the UV and X-rays have the same
origin, usually the jet component. From Table 1 we see that the
values of the model parameter $p$ for the power-law index of the
relativistic electrons are all in their ``reasonable'' range, and
that the relation between $\epsilon_e$ and $\epsilon_B$ follows the
trend $\epsilon_e \sim \sqrt{\epsilon_B}$ suggested by Medvedev
(2006). The values of other parameters are also ``typical.''
Exceptions are NGC~3368 and NGC~4736, where we have to adopt smaller
values of $\delta=0.05$ and 0.01, respectively. We note in this
context that large uncertainties exist in the value of $\delta$
(Sharma et al. 2007). The non-standard value of $\delta$ in these
two sources could also be because the values of some parameters such
as $s$ and $\alpha$ we fix in the model change from source to
source. {\em In summary, the overall good fitting results implies
that, the UV emission of all these sources generally can be well
fitted by the emission from the ADAF-jet model}. The one exception
is NGC~1052, whose predicted UV flux is much higher than observed
(Fig. 3). We note that the ratio of UV to X-ray flux of this source
is very small, actually the smallest in our sample. This source is
also very radio-loud, with a value for its radio-loudness parameter
as large as $\log R_{\rm UV} \approx 5$ (Maoz 2007).  The UV flux
adopted by Maoz et al. (2005) at 2500~\AA~ and 3300~\AA~ was
observed only once by {\it HST}\ in 2002, which was only roughly
half the flux observed in 1997 by Gabel et al. (2000). Maoz et al.
(2005) concluded that the factor of 2 variation was real, and that
this meant that the UV flux of the nucleus contains a significant
AGN contribution.

Yuan \& Cui (2005) predicted that the X-ray emission of the system
should be dominated by the jet rather than by the ADAF when the
X-ray luminosity in the 2--10 keV band, $L_{\rm 2-10 keV}$, is lower
than a critical value $L_{\rm X,crit}$: \be {\rm
log}\left(\frac{L_{\rm X,crit}}{L_{\rm Edd}}\right) =-5.356-0.17{\rm
log}\left(\frac{M}{\msun}\right). \label{eq:critlum} \ee This is
because the X-ray emission from the ADAF is roughly proportional to
$\dot{M}^2$, whereas the synchrotron emission from the jet is
proportional to $\dot{M}$ or $\dot{M}^{0.5}$.  Hence, as $\dot{M}$
decreases below a certain threshold, the synchrotron emission from
the jet will finally dominate the X-ray emission. This prediction
has been recently confirmed convincingly by modeling 16 LLAGNs with
good radio and X-ray spectra (Yuan, Yu \& Ho 2009). Here we can see
from Figures 1 and 2 that the current modeling results further
confirm it. Eight out of the 10 sources in our sample with hard
X-ray data\footnote{NGC~3652 has no data in the 2--10 keV band.} are
consistent with the prediction of Yuan \& Cui (2005). The exceptions
are M~81 and NGC~3998, which, despite having $L_{\rm 2-10keV} >
L_{\rm X,crit}$, are dominated by jets. However, as pointed out in
Yuan et al. (2009), the correlation is correct only in statistical
sense, and thus some outliers are not unexpected.

Regarding the spectrum in the UV, we note that for most sources,
because the data are quite heterogeneous and span a narrow range of
frequency, the uncertainty of the continuum shape is large. Given
this situation, we regard the overall match between data and model
to be satisfactory.

As a comparison, we have also tried to fit the UV spectrum of each
source with a standard thin disk model with the inner edge located
at the innermost stable circular orbit. We adjust the accretion rate
to fit the UV data. The results are shown by the dot-dashed-dashed
line in each panel of the figures; the parameters of the fit are
listed in Table 1. For most sources, the goodness of the fit is
similar to the ADAF model. Compared to the ADAF model, for NGC~4552
and especially M~87, the standard thin disk fit is clearly worse.
{\em By contrast, for M~81 and NGC~1052, the fit for the standard
thin disk actually looks better; the emission from the jet and ADAF
seems to overpredict the UV flux}. We note that this is also the
case for the pure jet model of Markoff et al. (2008) for M~81. The
UV flux falls far below the X-ray flux. Such a large X-ray/UV ratio
is also hard to be explained in the framework of the ``disk-corona''
scenario, as we explain in the following paragraph. One possible
culprit is inadequate correction for intrinsic absorption, although
no strong evidence for internal reddening is seen in the analysis of
the UV spectrum by Ho et al. (1996). Alternatively, the X-ray
emission of these two sources does not originate from a jet or
``standard'' ADAF, but from a ``modified'' ADAF in which there are
some cool clumps embedded in the hot flow. The black body radiation
of the cool clumps explains the origin of the UV, while the Compton
scattering of these UV photons produces the X-ray emission.

Although a standard thin disk can account for the UV emission in M~81 and
NGC~1052, it cannot, as is well known for AGNs in general, explain the X-ray
emission. The UV emission in M~81 represents $50\%$ the X-ray luminosity;
in NGC~1052, it is only $\sim 10\%$.  If we adopt a disk-corona model (e.g.,
Haardt \& Maraschi 1991), we have to require that most of the gravitational
energy be dissipated in the corona rather than in the disk. This is
the case for most sources in our sample. From Table 1 (Col. 15) we
can see that the UV luminosity is on average only $\sim 20\%$ of the
bolometric luminosity of each source. This is hard to understand
theoretically.  While most disk-corona models so far are
only phenomenological (e.g., Haardt \& Maraschi 1991; Svensson \&
Zdziarski 1994), recent three-dimensional radiation magneto-hydrodynamical
simulations of the vertical structure of a local patch of a standard thin
accretion disk shows that most of the dissipation occurs in the midplane
of the disk, with very little dissipation in the corona
(Hirose et al. 2006; Blaes 2007). This simulation result
may not be the final answer to the question, but it is in good
agreement with the observations of the high state of BH
X-ray binaries. The standard model for the high state is a standard thin
disk (Zdziarski \& Gierli\'nski 2004;  Done et al.  2007), which is
well fitted by a multi-temperature black body spectrum,
with very weak hard X-ray emission. On the other hand, even if enough
dissipation can occur in the corona to generate the requisite high
X-ray/UV ratios, the structure and emitted spectrum of the underlying
thin disk would be violently changed compared to the standard thin
disk, and it is unclear whether in that case the model can still
fit the UV spectrum.

\section{Summary and Discussion}

The accretion flow in LLAGNs is believed to be described by an ADAF,
which is different from the standard thin disk invoked for luminous
AGNs. Maoz (2007) disagrees.  Using new UV observations of a sample
of 13 LLAGNs, he shows that their X-ray/UV and radio/UV ratios are
similar to those of more powerful sources. Because of this, he
maintains that AGNs across all luminosities have essentally the same
accretion and radiative processes.

In this paper, we fit the SEDs of Maoz's sample using a model consisting of
an accretion flow described by central ADAF within a truncation radius
$R_{\rm tr}$ plus a thin disk outside of this radius. In the innermost region
a jet is included. We find that for almost all sources in our sample, our
model can give a good fit to the SED ranging from radio, optical/UV, to
X-rays, without the need of a standard thin disk extending to the innermost
stable circular orbit. We would like to
emphasize that the free parameters in our model are almost
completely fixed by the radio and X-ray data. In other words, we
have almost no room to adjust model parameters to fit the UV data.
This result strongly demonstrates that the apparent similarity of the
SEDs of low- and high-luminosity AGNs cannot rule out the ADAF model.

We have also tried to fit the UV data using the standard thin disk
model alone. Compared to the ADAF fitting result, the goodness of
the standard thin disk model fitting is similar; two sources seem to
be better fitted, while two are worse. As is well known, the
standard thin disk model can only fit the UV. To explain other data,
such as the radio and X-rays, in the framework of the standard thin
disk, two additional components, namely a jet and a corona, are
required. This situation is similar to the ADAF model where three
components are also required. The ADAF model is superior to the thin
disk scenario because of the following reasons. First, the dynamics
of an ADAF are better understood than those in the disk-corona
model. The latter contains many more theoretical uncertainties, such
as the main heating mechanism of the corona and the physics of
magnetic reconnection. Second, theoretical studies indicate that a
jet --- a near-ubiquitous component observed in LLAGNs and X-ray
binaries in their low-state --- arises much more naturally in an
ADAF than in a thin disk. Third, although it is in principle
possible to produce the X-ray spectrum by a disk-corona model, it is
questionable whether enough gravitational energy can be dissipated
in the corona to account for the typically low UV/X-ray flux ratios
that characterize LLAGNs. Currently the most serious attempts to
address this question comes from three-dimensional
magneto-hydrodynamical simulations of a patchy disk. The result
shows that little dissipation occurs in the corona. On the other
hand, even if most of the energy were dissipated in the corona the
emitted spectrum from the thin disk in this case would be quite
different from that of a pure standard thin disk, and it is
questionable whether such a ``modified'' thin disk can still explain
the UV spectrum. In fact, observations of the high/soft state of
black hole X-ray binaries, which are widely believed to be powered
by a standard thin disk, support the above two arguments, namely
that the thin disk launches no jet and produces extremely weak X-ray
emission. Lastly, the SED is only one of several characteristics of
LLAGNs that needs to be explained. As discussed in the Introduction,
LLAGNs possess a set of other distinctive properties, and any
successful model should try to simultaneously account for them all.
As outlined in Ho (2008), a three-component central engine
consisting of an ADAF, a truncated thin disk, and a jet provides a
promising, observationally and physically well-motivated framework
for understanding LLAGNs.

We propose a set of diagnostics that can distinguish the origin
of the UV emission.  1) Since the UV radiation in the ADAF model comes from
the combination of an inner hot flow, a truncated thin disk,
and a jet, in general we do not expect a blackbody-like bump
in the optical/UV band, as predicted by a standard thin disk. So it will be
useful to have reliable, broader wavelength coverage across the SED,
especially over the near-infrared, optical, and UV bands. The current
spectral range is too narrow and the data too heterogeneous to
constrain models. 2) The emission from the standard thin disk is
very steady. This is evidenced by the very weak variability of the
soft state of BH X-ray binaries in which a thin disk is
believed to be present. By contrast, the
variability from an ADAF is expected to be stronger (Manmoto et al.
1996; Igumenshchev \& Abramowicz 1999).  So we predict that UV
variability should be significant and correlated with X-ray
emission. 3) When the source is very dim, with $L_{\rm X}\la L_{\rm
crit}$ as defined in Equation (1), the UV spectrum is dominated by the
synchrotron emission of the jet. In this case, we expect the UV
spectrum to be similar to that in the X-rays. If described by a
power-law, $F_{\nu}\propto \nu^{\alpha}$, we expect
$\alpha\approx -1$ if the relativistic electrons in the jet
are accelerated by shocks. 4) In this case, we also expect a high level of
linear polarization in the UV band; the magnitude and angle of the
polarization should be similar to those in the X-ray band.

\begin{acknowledgements}

We thank the referee, Dr. Dan Maoz, for constructive criticisms.
This work was supported in part by the Natural Science Foundation of
China (grants 10773024, 10833002, 10821302, and 10825314), the
National Basic Research Program of China (973 Program 2009CB824800),
and the CAS/SAFEA International Partnership Program for Creative
Research Teams.

\end{acknowledgements}


\begin{deluxetable}{lllllllllllllllllllllllllllllll}
\rotate \tablewidth{24.cm} \tablecaption{Properties of the Sample
} \tabletypesize{\scriptsize} 
\tablehead{Object  &  $M_{\rm BH}(\msun)$ & $R_{\rm tr} (R_{\rm S})$ & $L_{\rm 2-10keV} (L_{\rm Edd})$ & $L_{\rm X,crit} (L_{\rm Edd})$ & Ref. & $\dot{m}_{\rm out}(2000 r_{\rm g})$  & $\dot{m}_{\rm jet}$ & $\Gamma_{\rm jet}$ &$\theta$ ($^{\circ}$)&$\epsilon_e$ &$\epsilon_B$ &$p$ & $\dot{m}_{\rm SSD}$ & log$(L_{\rm SSD} / L_{\rm bol}) $\\
(1) & (2) & (3) & (4) & (5) & (6) & (7) & (8) & (9) & (10) & (11) &
(12) & (13) & (14) & (15) & & }

\startdata
M~81 & $7. \times 10^7$ & 100. & $2.3 \times 10^{-6}$ & $2. \times 10^{-7}$ & 5,6,8,11 & $1. \times 10^{-4}$ &$2.4 \times 10^{-7}$ & 10& 25 &0.25&0.02&1.8& $1. \times 10^{-4}$ & -0.54 &\\
M~87 & $3.4 \times 10^9$ & $10^{4}$ & $8.1 \times 10^{-8}$ & $1. \times 10^{-7}$ & 1,5,13 & $7. \times 10^{-5}$ & $1. \times 10^{-8}$ & 70& 19 &0.2&0.02&2.5& $1. \times 10^{-5}$ & -0.77 &\\
NGC~1052 & $1.26 \times 10^8$ & 75. & $5.9 \times 10^{-6}$ & $1.85 \times 10^{-7}$ & 5 & $1.3 \times 10^{-3}$ & $8. \times 10^{-5}$ & 10& 60 &0.2&0.02&2.3& $6. \times 10^{-5}$ & -1.37 &\\
NGC~3368 & $2.5 \times 10^7$ & 40. & $6.3 \times 10^{-7}$ & $2.47 \times 10^{-7}$ & 5 & $3.2 \times 10^{-3}$ & $8. \times 10^{-6}$ & 10& 60 &0.1&0.01&2.3& $7. \times 10^{-5}$ & -0.36 &\\
NGC~3642 & $1.26 \times 10^7$ & 60. & ... & $1.85 \times 10^{-7}$ & 4,5 & $1. \times 10^{-2}$ & $3. \times 10^{-6}$ & 10& 60 &0.2&0.02&2.2& $1. \times 10^{-3}$ & -0.94 &\\
NGC~3998 & $7. \times 10^8$ & 100. & $3. \times 10^{-6}$ & $1.4 \times 10^{-7}$ & 5,7,10 & $2. \times 10^{-4}$ & $9.8 \times 10^{-6}$ & 10& 20 &0.2&0.02&1.8& $5. \times 10^{-5}$ & -0.94 &\\
NGC~4203 & $1. \times 10^7$ & 250. & $1.85 \times 10^{-5}$ & $2.84 \times 10^{-7}$ & 5,12 & $5.5 \times 10^{-3}$ & $1.5 \times 10^{-6}$ & 10& 25 &0.1&0.02&2.2& $5. \times 10^{-4}$ & -1.04 &\\
NGC~4552 & $3.16 \times 10^8$ & 300. & $4.3 \times 10^{-8}$ & $1.6 \times 10^{-7}$ & 5 & $4. \times 10^{-5}$ & $3.5 \times 10^{-6}$ & 10& 60 &0.15&0.01&2.1& $1. \times 10^{-5}$ & -0.04 &\\
NGC~4579 & $4. \times 10^6$ & 50. & $3. \times 10^{-4}$ & $3.3 \times 10^{-7}$ & 2,5,12 & $2.3 \times 10^{-2}$ & $8. \times 10^{-5}$ & 10& 60 &0.2&0.01&2.3& $9. \times 10^{-3}$ & -0.59 &\\
NGC~4594 & $1. \times 10^9$ & 300. & $1.2 \times 10^{-7}$ & $1.3 \times 10^{-7}$ & 2,3,5,9& $3. \times 10^{-5}$ & $4. \times 10^{-7}$ & 10& 45 &0.3&0.02&1.8& $2. \times 10^{-5}$ & -0.16 &\\
NGC~4736 & $1.26 \times 10^7$ & 75. & $5. \times 10^{-7}$ & $2.9 \times 10^{-7}$ & 5& $5.2 \times 10^{-3}$ & $1.5 \times 10^{-6}$ & 10& 60 &0.1&0.01&2.1& $6. \times 10^{-5}$ & -0.37 &\\

\enddata
\tablecomments{ Col. (1): Name of object. Col. (2): Mass of the BH.
Col. (3): Truncation radius. Col. (4): Luminosity in the 2--10 keV
band. Col. (5) The critical luminosity from Equation (1). Col. (6):
References from which part of the observational data come from. Col.
(7): Mass accretion rate extrapolated from $R {\rm tr}=2000\,r_g$.
Col. (8): Mass lost rate in the jet. Col. (9): Lorentz factor of the
jet. Col. (10): Viewing angle of the jet. Col. (11): Fraction of
shock energy entering into electrons. Col. (12): Fraction of shock
energy entering into magnetic field. Col. (13): Spectral index of
power-law electrons. Col. (14): Accretion rate of the standard thin
disk extending to the innermost stable circular orbit. Col. (15):
Ratio between the luminosity of the standard thin disk and the
bolometric luminosity.}

\tablerefs{ (1) Evans et al. 2006; (2) Ho 1999; (3) Hummel et al.
1984; (4) Komossa et al. 1999; (5) Maoz 2007; (6) Markoff et al.
2008; (7) Pellegrini et al. 2000b; (8) Pellegrini et al. 2000a; (9)
Pellegrini et al. 2003; (10) Ptak et al. 2004; (11) Sch\"odel et al.
2007; (12) Terashima et al. 2002; (13) Wilson \& Yang 2002. }
\end{deluxetable}


{}

\clearpage

\begin{figure}
\epsscale{0.45} \plotone{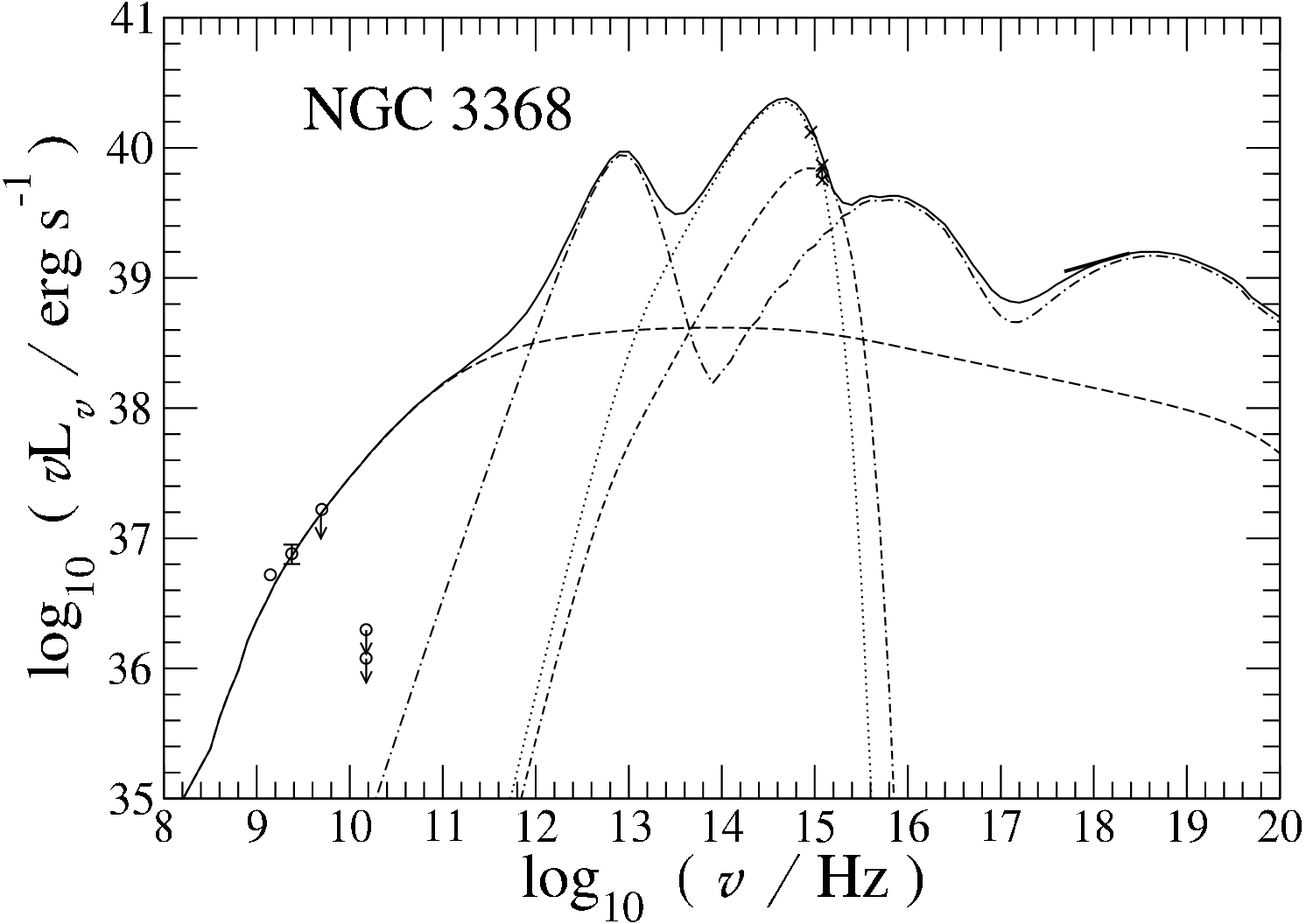}\hspace{1.cm} \epsscale{0.45}
\plotone{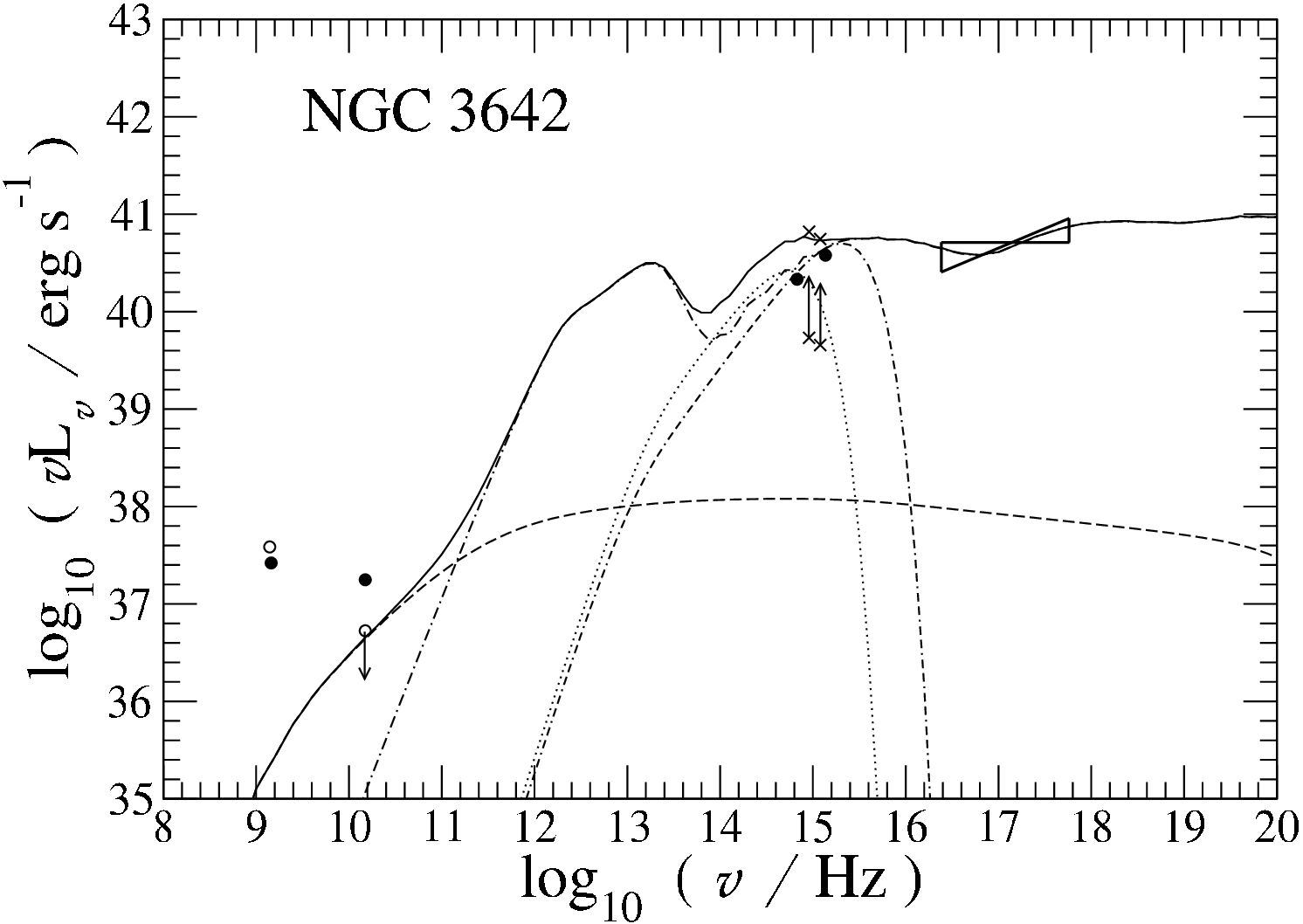} \vspace{0.2in}\epsscale{0.45} \plotone{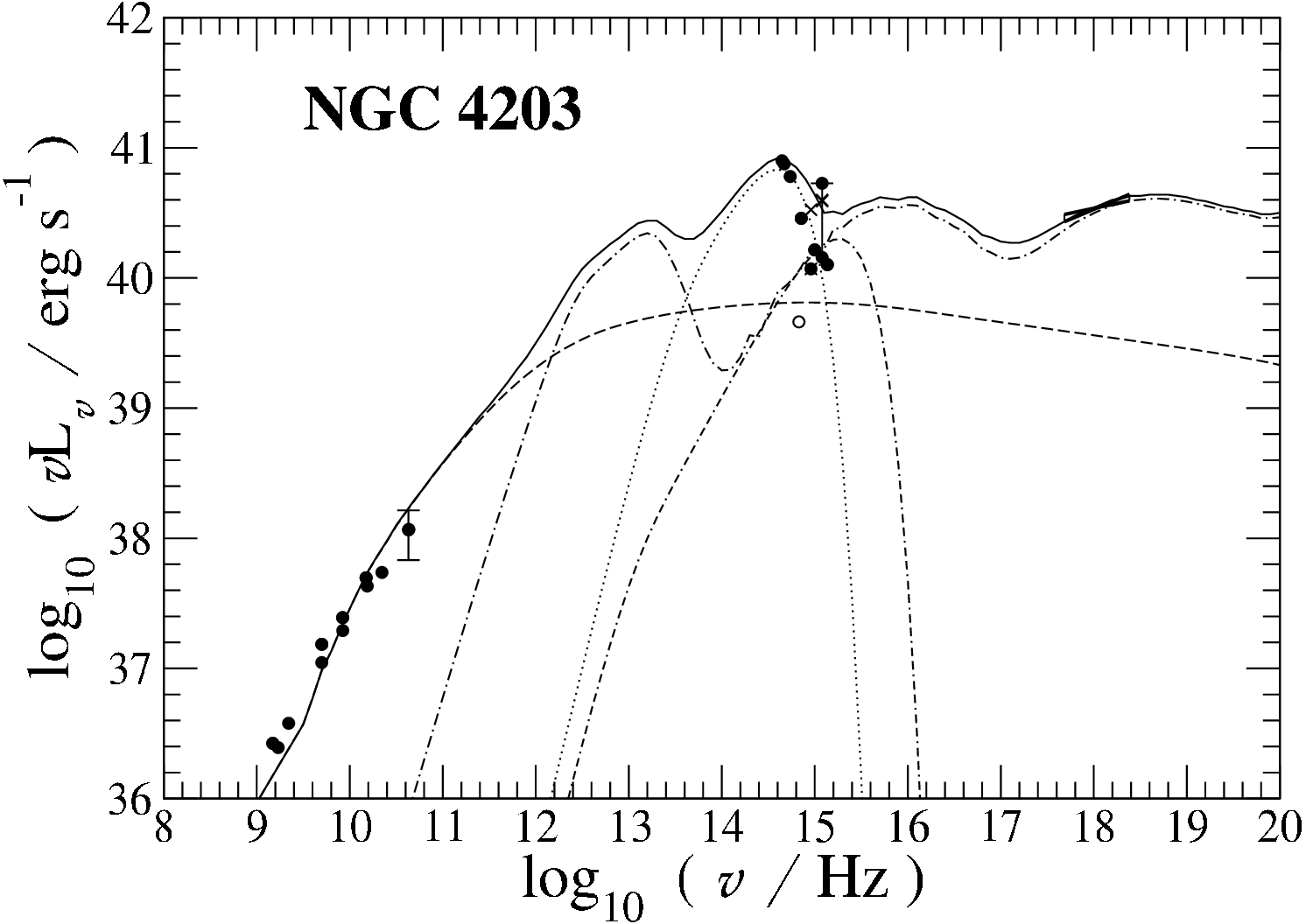}
\hspace{1cm} \epsscale{0.45} \plotone{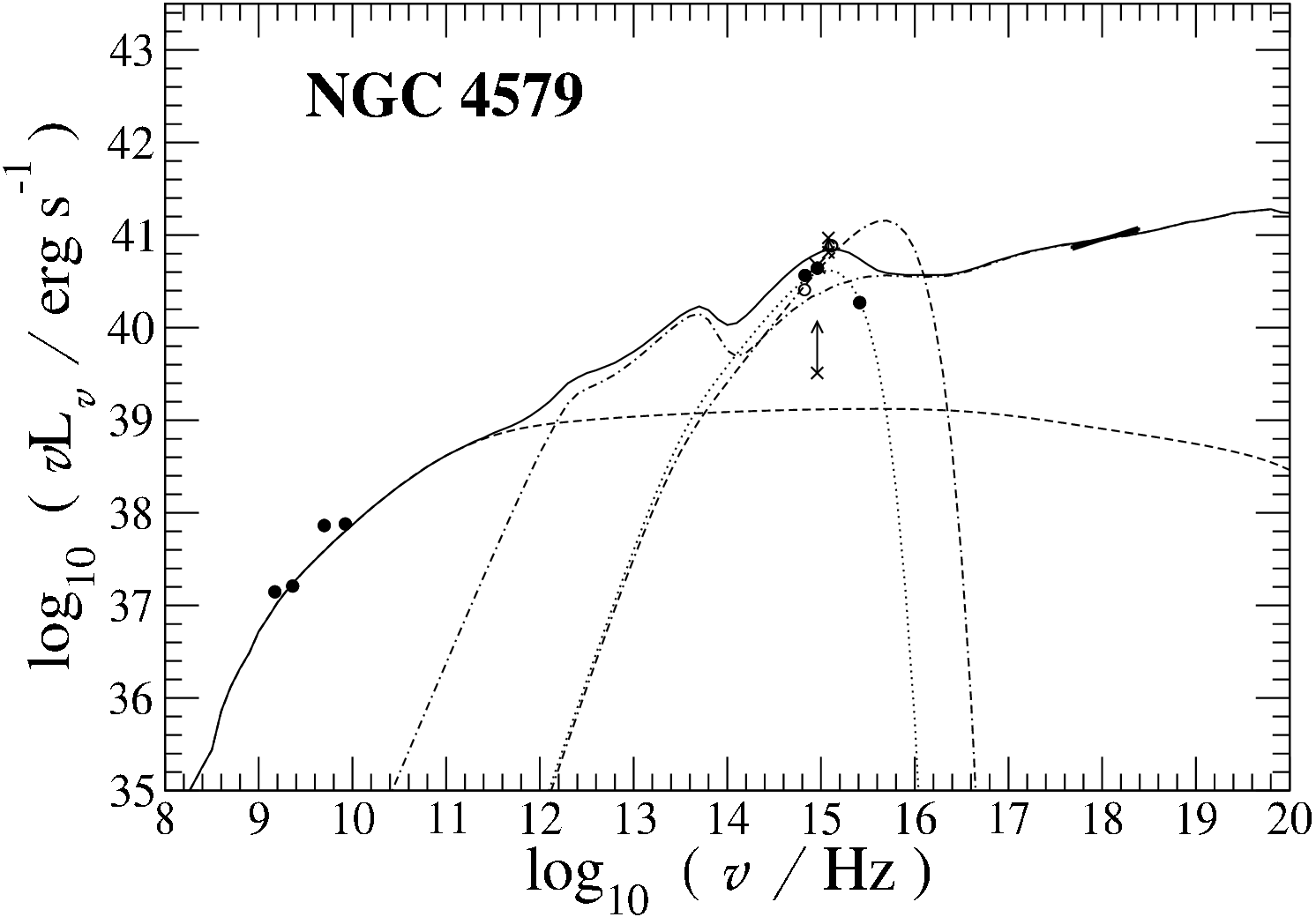}\vspace{0.2in}
\epsscale{0.45} \plotone{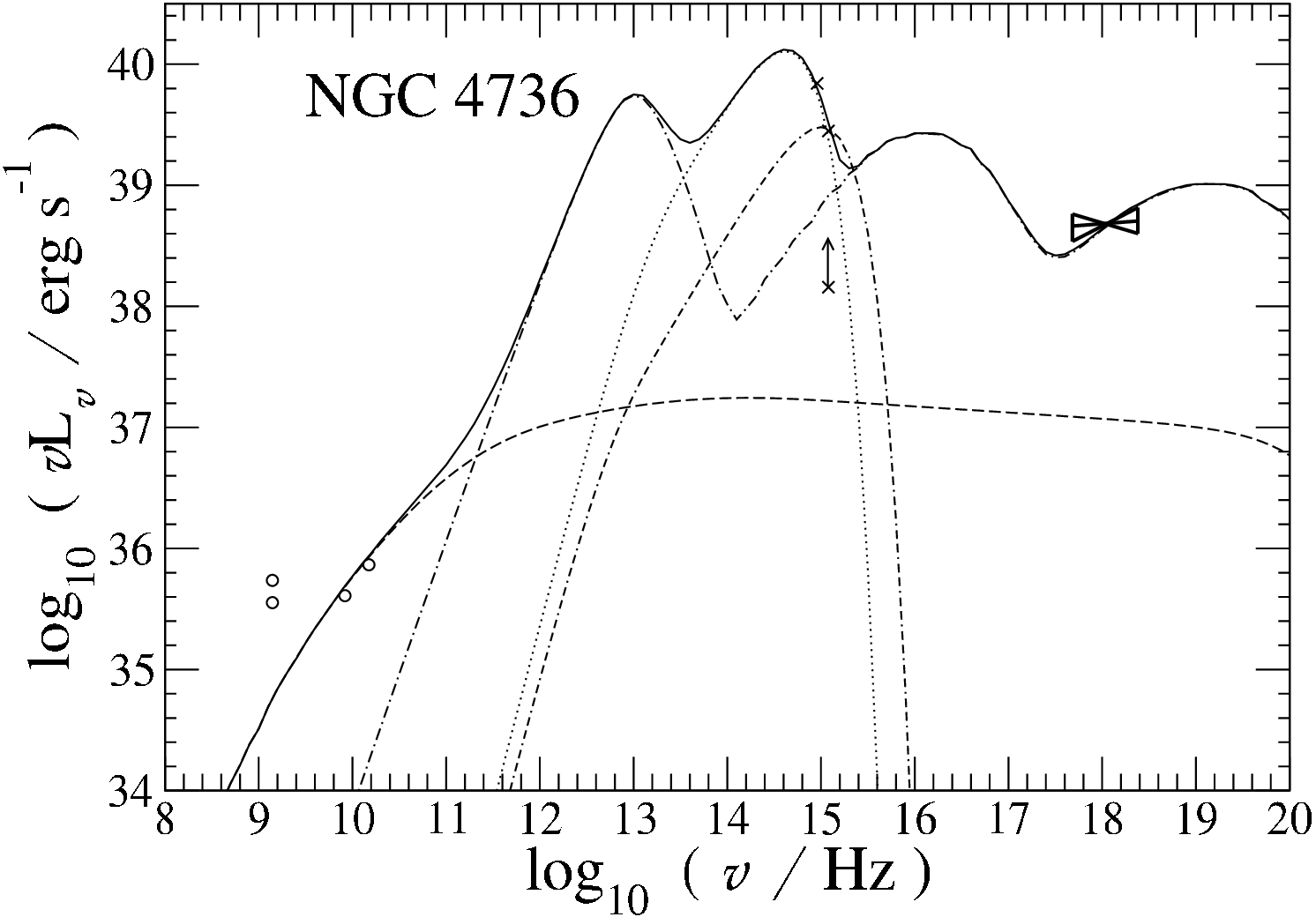}\vspace{0.2in} \caption{The
ADAF-dominated sources. The dot-dashed, dashed, dotted, and solid
lines show the emitted spectrum from the ADAF, the jet, the
truncated standard thin disk, and their sum, respectively, The
dot-dashed-dashed line shows the emission from a standard thin disk
extending to the innermost stable circular orbit. Reliable data
points are plotted as solid symbols, whereas points severely
affected by host galaxy contamination or extinction are plotted as
open symbols. The UV data from Maoz (2007) are marked as crosses.}
\end{figure}

\clearpage

\begin{figure}
\epsscale{0.45} \plotone{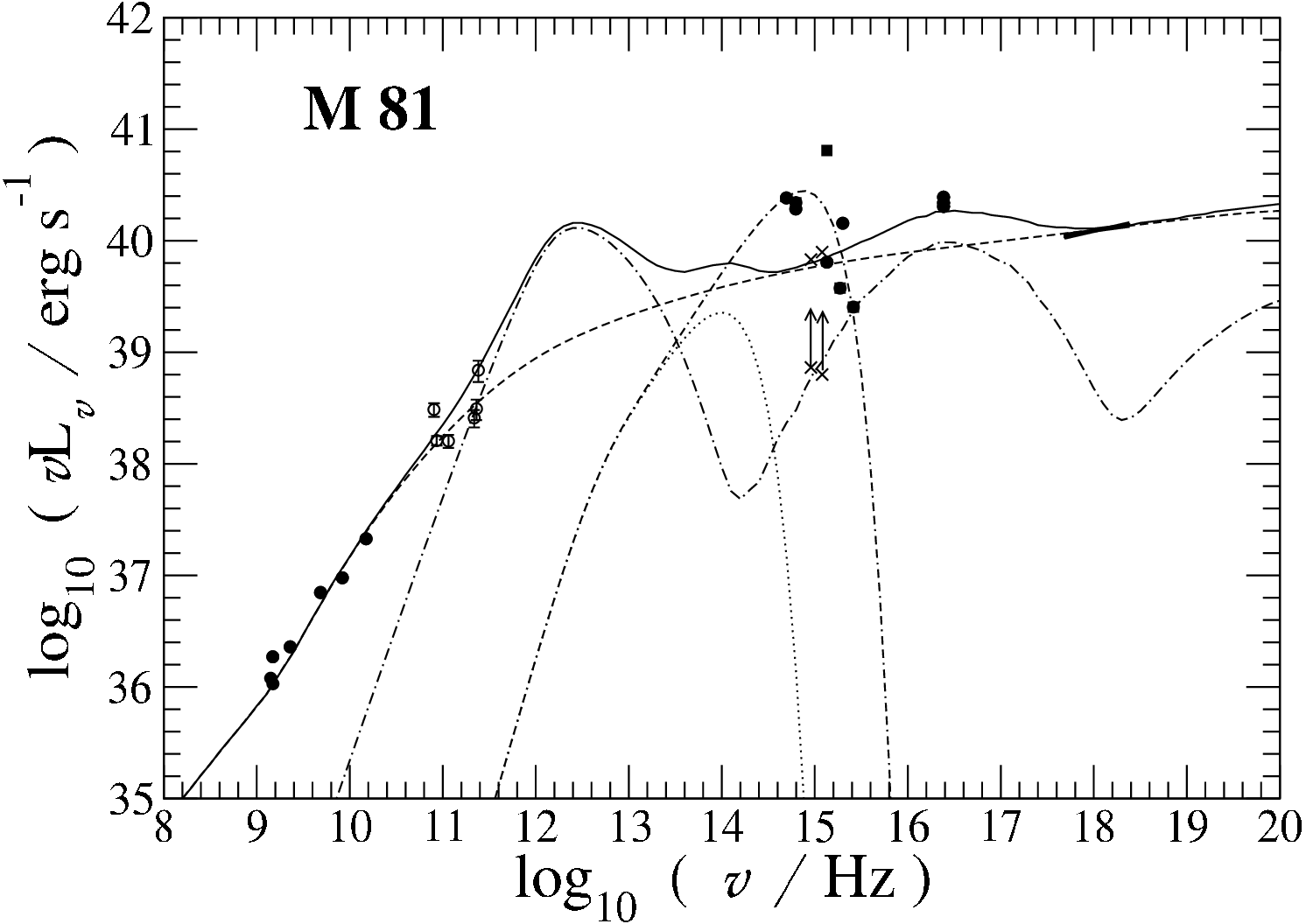}\hspace{1cm} \epsscale{0.45}
\plotone{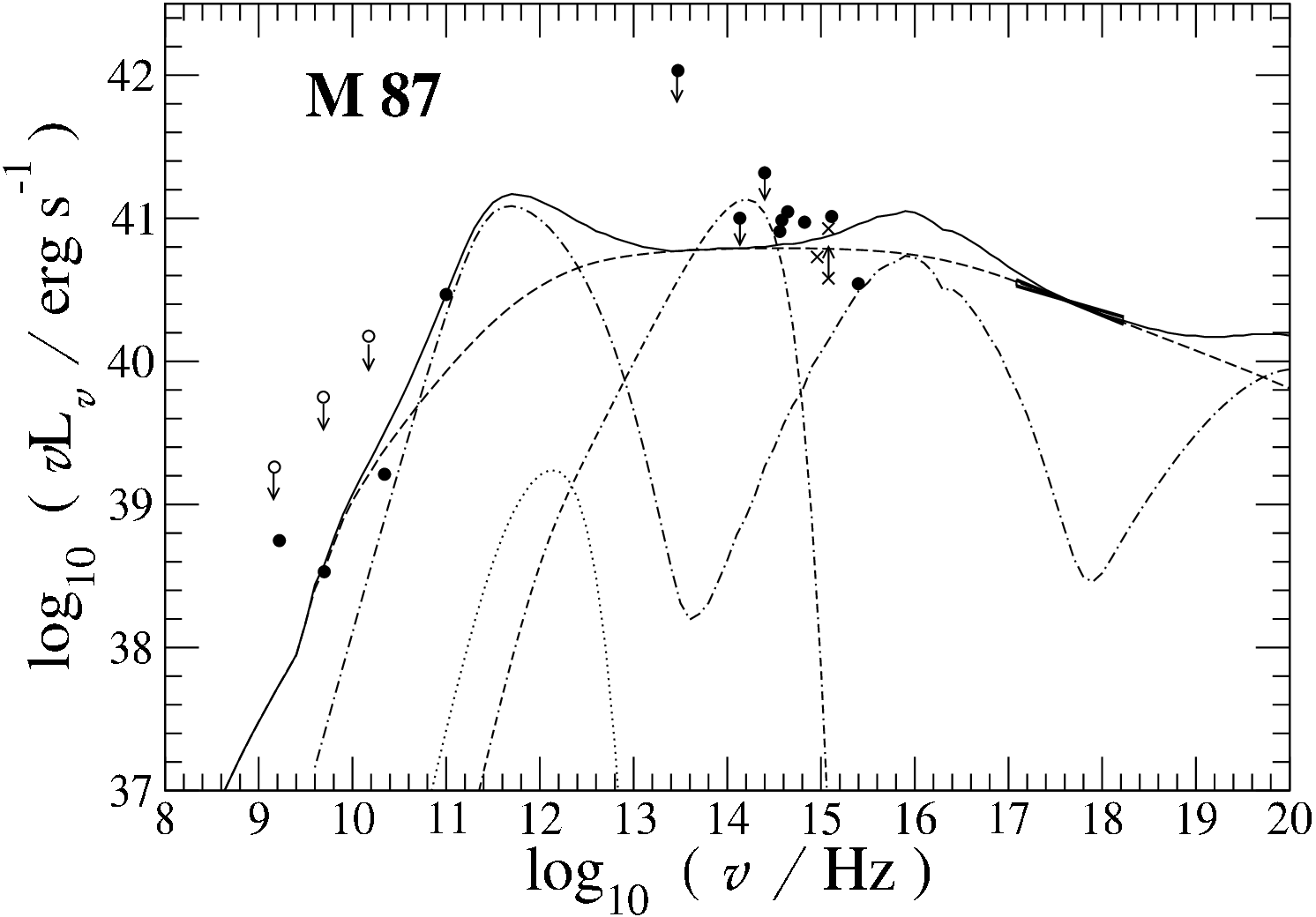} \vspace{0.2in} \epsscale{0.45}
\plotone{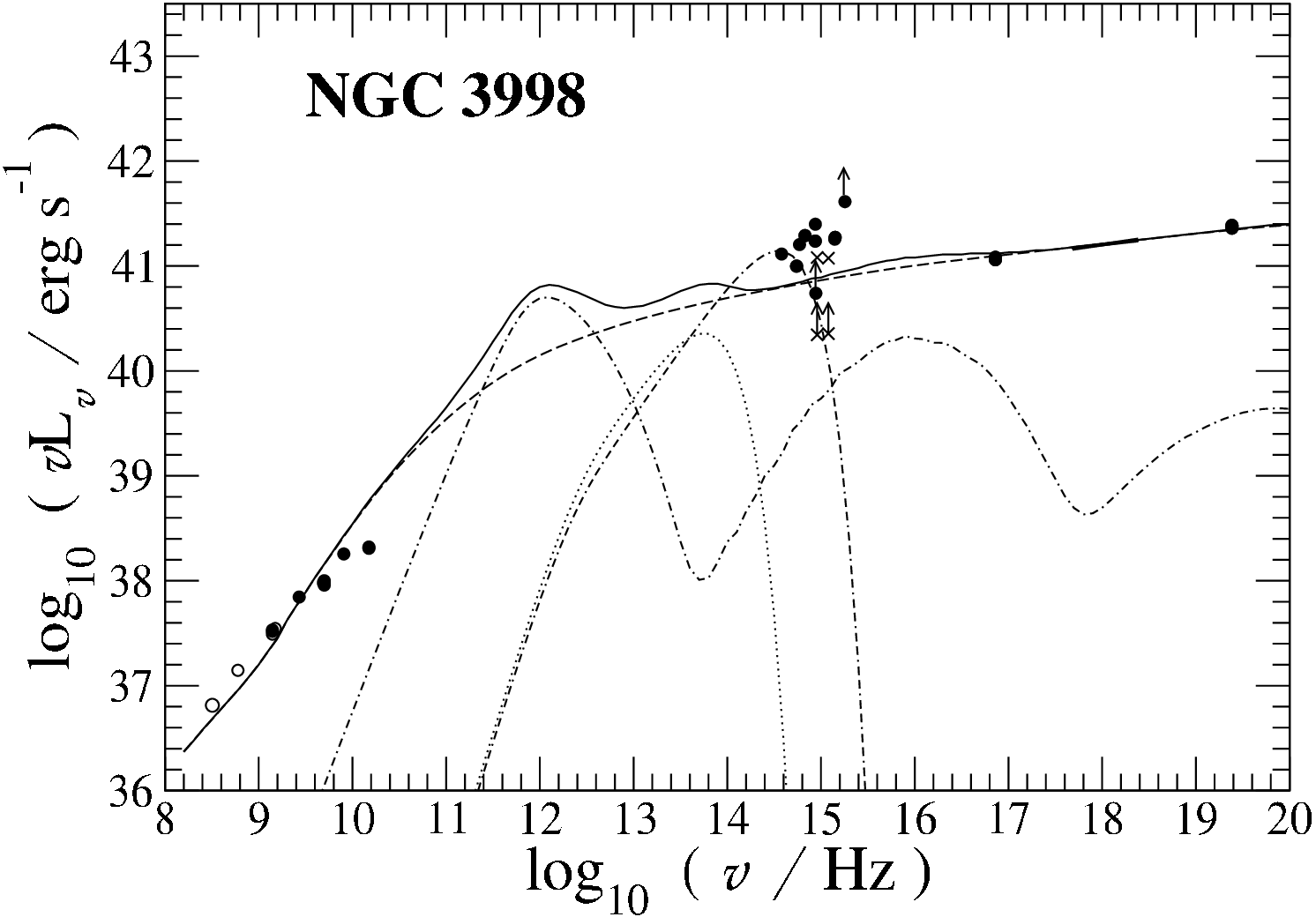}\hspace{1cm} \epsscale{0.45}
\plotone{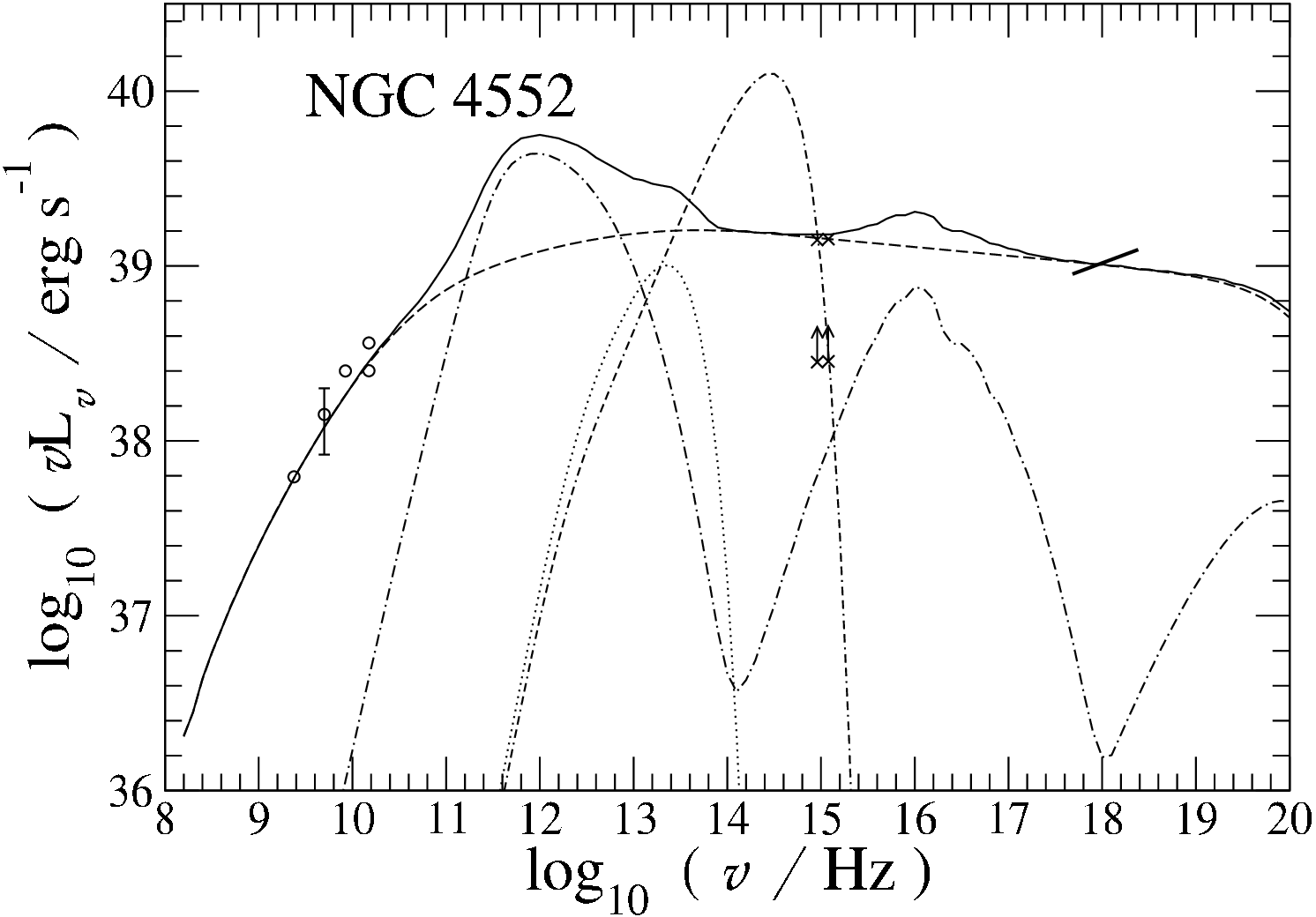}\vspace{0.2in} \epsscale{0.45}
\plotone{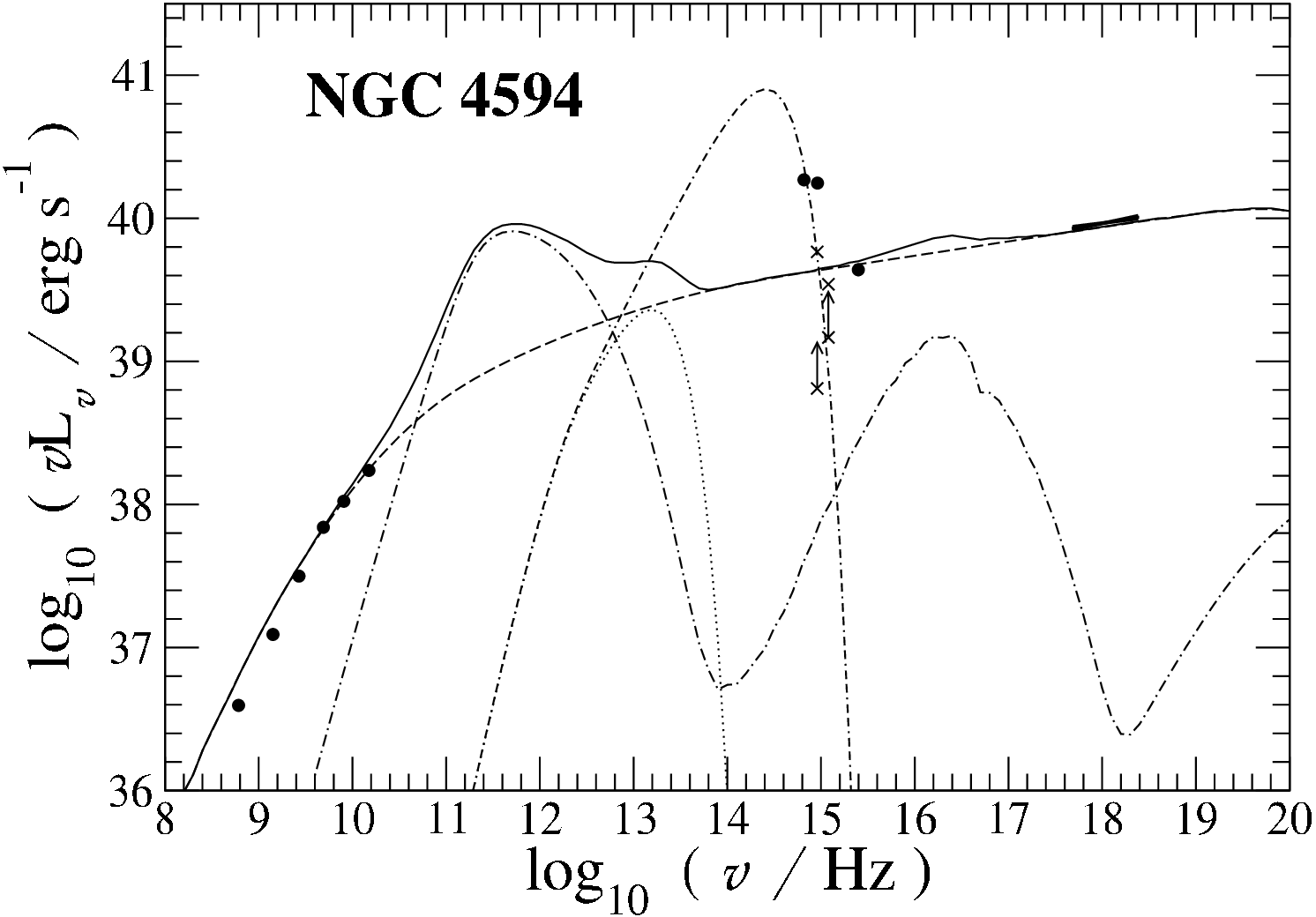}\vspace{0.2in} \caption{The jet-dominated sources.
The dot-dashed, dashed, dotted, and solid lines show the emitted
spectrum from the ADAF, the jet, the truncated standard thin disk,
and their sum, respectively, The dot-dashed-dashed line shows the
emission from a standard thin disk extending to the innermost stable
circular orbit. Reliable data points are plotted as solid symbols,
whereas points severely affected by host galaxy contamination or
extinction are plotted as open symbols. The UV data from Maoz (2007)
are marked as crosses.}

\end{figure}

\clearpage

\begin{figure}
\epsscale{0.75} \plotone{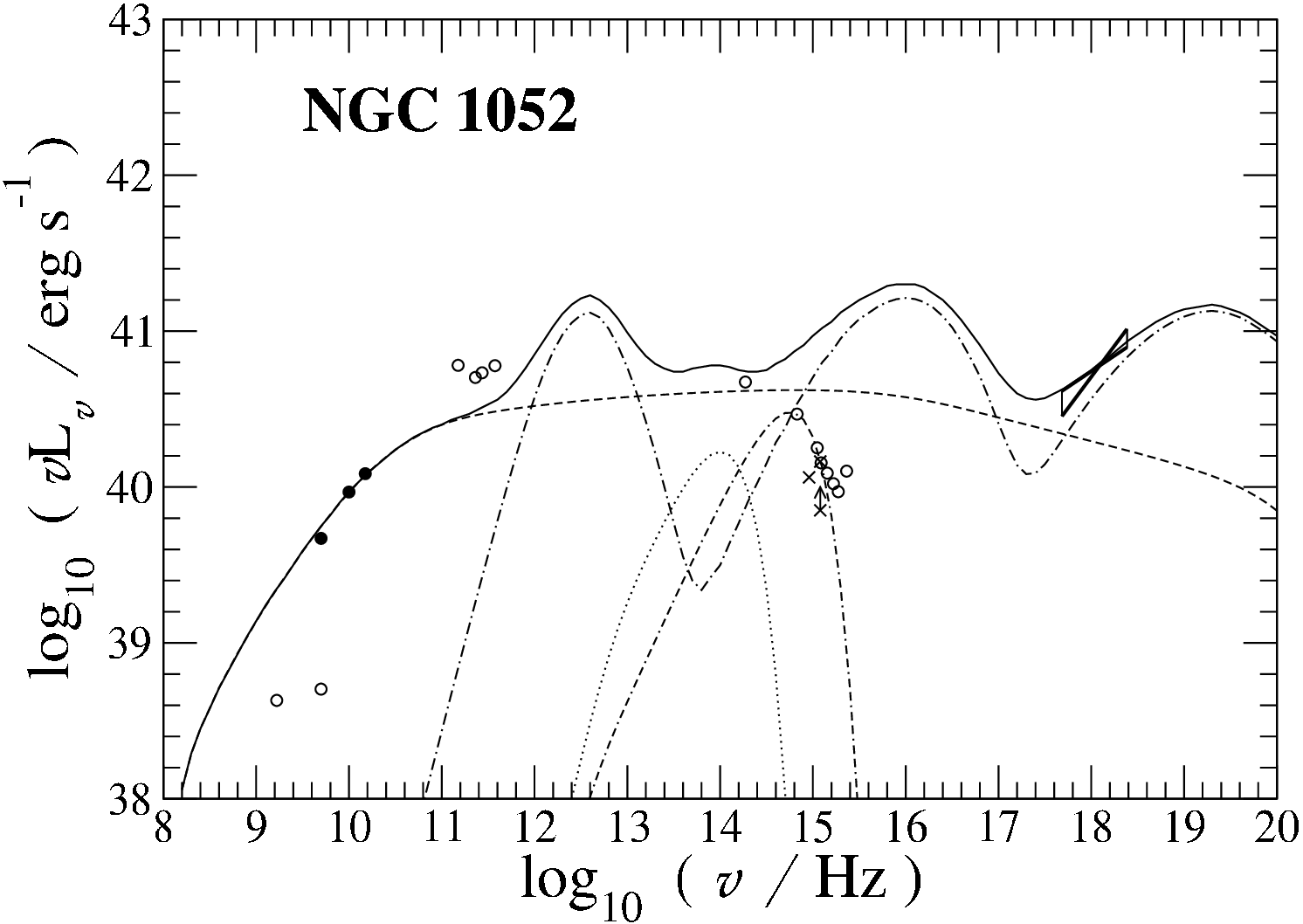}\hspace{1cm}  \caption{The special
case of NGC~1052.  The dot-dashed, dashed, dotted, and solid lines
show the emitted spectrum from the ADAF, the jet, the truncated
standard thin disk, and their sum, respectively, The
dot-dashed-dashed line shows the emission from a standard thin disk
extending to the innermost stable circular orbit. Reliable data
points are plotted as solid symbols, whereas points severely
affected by host galaxy contamination or extinction are plotted as
open symbols. The UV data from Maoz (2007) are marked as crosses.}

\end{figure}

\end{document}